# Direct observation of individual charges and their dynamics on graphene by low-energy electron holography

Tatiana Latychevskaia*, Flavio Wicki, Jean-Nicolas Longchamp, Conrad Escher, Hans-Werner Fink

Physics Department, University of Zurich,

Winterthurerstrasse 190, 8057 Zurich, Switzerland

*Corresponding author: tatiana@physik.uzh.ch

## ABSTRACT

Visualizing individual charges confined to molecules and observing their dynamics with high spatial resolution is a challenge for advancing various fields in science, ranging from mesoscopic physics to electron transfer events in biological molecules. We show here, that the high sensitivity of low-energy electrons to local electric fields can be employed to directly visualize individual charged adsorbates and to study their behaviour in a quantitative way. This makes electron holography a unique probing tool for directly visualising charge distributions with a sensitivity of a fraction of an elementary charge. Moreover, spatial resolution in the nanometer range and fast data acquisition inherent to lens-less low-energy electron holography allows for direct visual inspection of charge transfer processes.

## Contents

# 1. Introduction

There exist just a few tools, most of them based on scanning probe technologies that allow for indirect imaging of individual charges[1-3]. Recently, Gatel et al demonstrated that high energy off-axis electron holography can be employed for imaging individual charges by applying contour integration data analysis[4]. We show here that low-energy electron holography can be employed for the direct visualization of individual charges bound to adsorbates on graphene. Graphene is partly transparent to low-energy electrons (30 – 250 eV)[5-6] and can be used as a support for samples[7] to be studied using Low-Energy Electron Point Source (LEEPS) microscopy[8]. Depending on the graphene preparation method, adsorbates of different chemical specificity are present. These individual adsorbates might be light organic molecules stemming from residual solvents[9] or from contamination due to air exposure[10]; possibly also individual C, H or Si atoms[11-12], other possible candidates are metal atoms[13]. Previous studies indicate that there is a charge transfer between graphene and its adsorbates[13]. Depending on the charge transfer direction, the adsorbates represent either a highly localized negatively or positively charged entity on graphene. Density functional theory (DFT) calculations show that almost all metal adatoms transfer electrons to graphene, whereby the transferred charge $\Delta q$ is typically $-1.6e < \Delta q < 0$, where $e$ is the elementary charge[13], except for Au adatoms for which $\Delta q = +0.18e$ [14]. Accordingly, most individual metal atoms may constitute highly localized charges on graphene. Furthermore, DFT calculations predict that charge transfer between graphene and adsorbed small molecules can lead to both, negatively or positively charged entities: $\Delta q = -0.025e$ for $H_2O$, $\Delta q = -0.099e$ for $NO_2$, $\Delta q = 0.012e$ for CO, $\Delta q = 0.018e$ for NO, and $\Delta q = 0.027e$ for $NH_3$[15]. Electrons of low kinetic energy are sensitive to local variations in the electric potential distribution confined to individual adsorbates on freestanding graphene. We show how individual charges of the order of just one elementary charge associated with the adsorbates on graphene can directly be visualized by low-energy electron holography. And as a consequence also charge transfer processes as well as the diffusion of adsorbates too small for direct observation can be studied.

## 2. Experimental arrangement

The low-energy electron holographic experimental setup has previously been described in the literature[8] and is shown in Figure 1a. A hologram is formed in the detector plane as a result of interference between the wave scattered by the object, and the unscattered (reference) wave[16-17]. The sample can be numerically reconstructed from such a hologram by propagation of the wavefront from the detector plane backwards to the object plane[18]. Ultraclean freestanding graphene spanning holes of 2 micrometers in diameter in a Pd/Cr covered silicon nitride membrane is prepared following the procedure described elsewhere[19]. However, if the sample is not transferred into the vacuum chamber fast enough, occasionally small, individual adsorbates are found when investigating freestanding graphene by LEEPS microscopy. Low-energy electron holograms of adsorbates on freestanding graphene are shown in Figure 1b – c, where a distribution of dark and bright features is apparent. Bright spots, as apparent in Figure 1b – c, are usually observed even when the graphene samples are carefully prepared. When applying a conventional hologram reconstruction routine[18, 20-21] to the pattern shown in Figure 1c, the dark features converge and reveal well resolved clusters of adsorbates as small as 2 nm in diameter, as shown in Figure 1d. In the course of the reconstruction procedure, the object distribution at different source-to-sample distances is calculated and converges towards an in-focus object reconstruction; beyond that distance the object distribution is diverging again. The source-to-sample distance is derived from the position at which the reconstructed objects, as for example clusters, appear in focus. The resolution of the reconstructed objects is determined by a Fourier transformation of the hologram respectively the reconstruction[22], details are provided in the Supporting Information. The concentric interference fringes around the reconstructed objects are due to the so-called twin-image effect which is intrinsic to in-line holography[17, 23]. In contrast to the holograms of the clusters in Figure 1c – d, the bright spots do not lead to a meaningful object reconstruction but preserve the same blurry appearance at all reconstruction distances. We refer to such occurrences in holograms that result in a non-meaningful object reconstruction as "spots" throughout the manuscript.

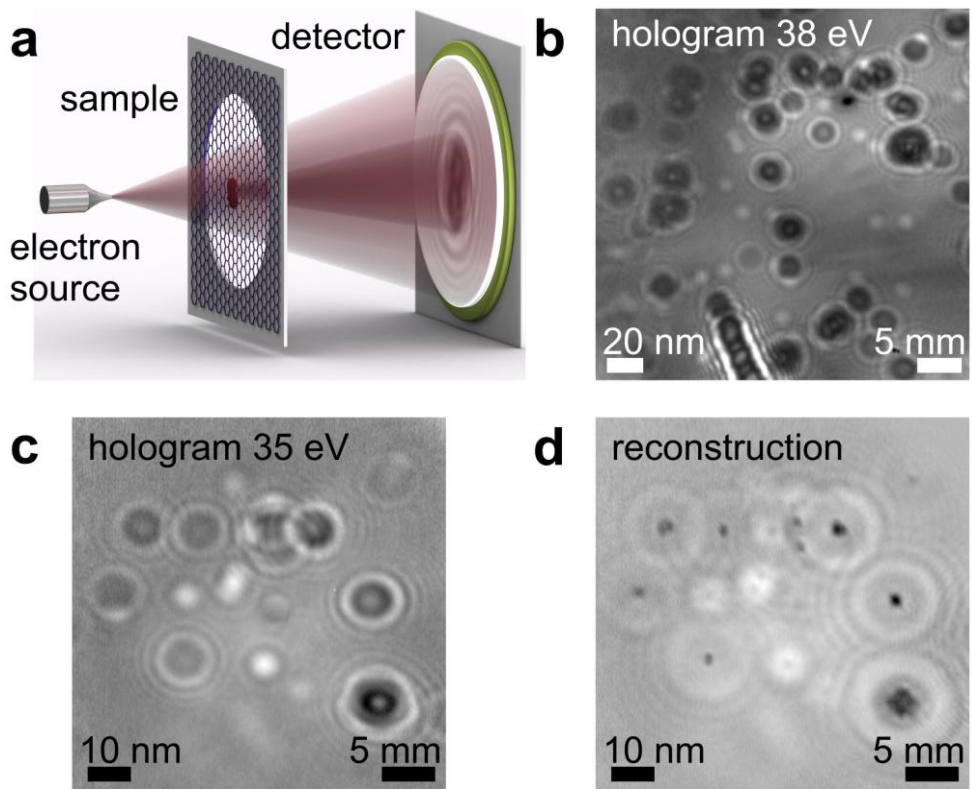

**Figure 1.** Imaging of adsorbates on freestanding graphene by low-energy electron holography. (a) Experimental scheme. (b) Low-energy electron hologram of adsorbates on graphene acquired with electrons of 38 eV at a tip-to-sample distance of 230 nm (Movie S1), at a resolution of 1.0 nm. (c) Another low-energy electron hologram of a graphene sample acquired with electrons of 35 eV at a tip-to-sample distance of 116 nm, at a resolution of 0.8 nm. (d) Amplitude reconstruction of the hologram in (c). The source-to-detector distance corresponding to the holograms shown in (b) and (c) is 47 mm. The scale bars indicate the sizes in the object plane (left) and in the detector plane (right).

## 3. Holograms of individual charges

Figure 2a depicts a normalized hologram[20, 23] where four bright features are observed. The corresponding radial distributions, plotted in Figure 2b, show that all four intensity distributions display the first and the second minima at approximately the same radial coordinates. In order to verify the possible origin of the bright features we performed a series of simulations: a hole, a single atom adsorbate and a charged adsorbate, as discussed in the Supporting Information. The simulated holograms of a hole, as well as those of neutral adatoms do not match the experimentally observed holograms, see Figure S3. Simulated holograms of charged adsorbates are shown in Figure 2c, the corresponding radial distributions are plotted in Figure 2e and an illustration of the arrangement of a charged adsorbate on the graphene surface is provided in Figure 2d. The holograms of $(-e)$ and $(+e)$ exhibit reversed contrast. This distinction is preserved at all studied electron energies: a positive charge results in a bright spot and a negative charge results in a dark spot. Additional simulations are presented in the Supporting Information and Figure S4. All simulated holograms of point charges exhibit the same diameter of the zero-diffraction order and approximately the same position of the minima, in good agreement with the experimental observation, as evident when comparing Figure 2e with 2b. In low-energy electron holography, a charged object not only influences the object wave, but also distorts the reference wave, which complicates the interpretation of the object reconstruction[21, 24-25]. Thus, the fact that the reconstructions of bright features do not converge to meaningful objects can be attributed to the electron wave being diffracted by charged objects. It was also verified that the simulated hologram of a charge leads to a non-meaningful reconstruction. We thus attribute the observed spots in the holograms to small adsorbates carrying positive or negative charges.

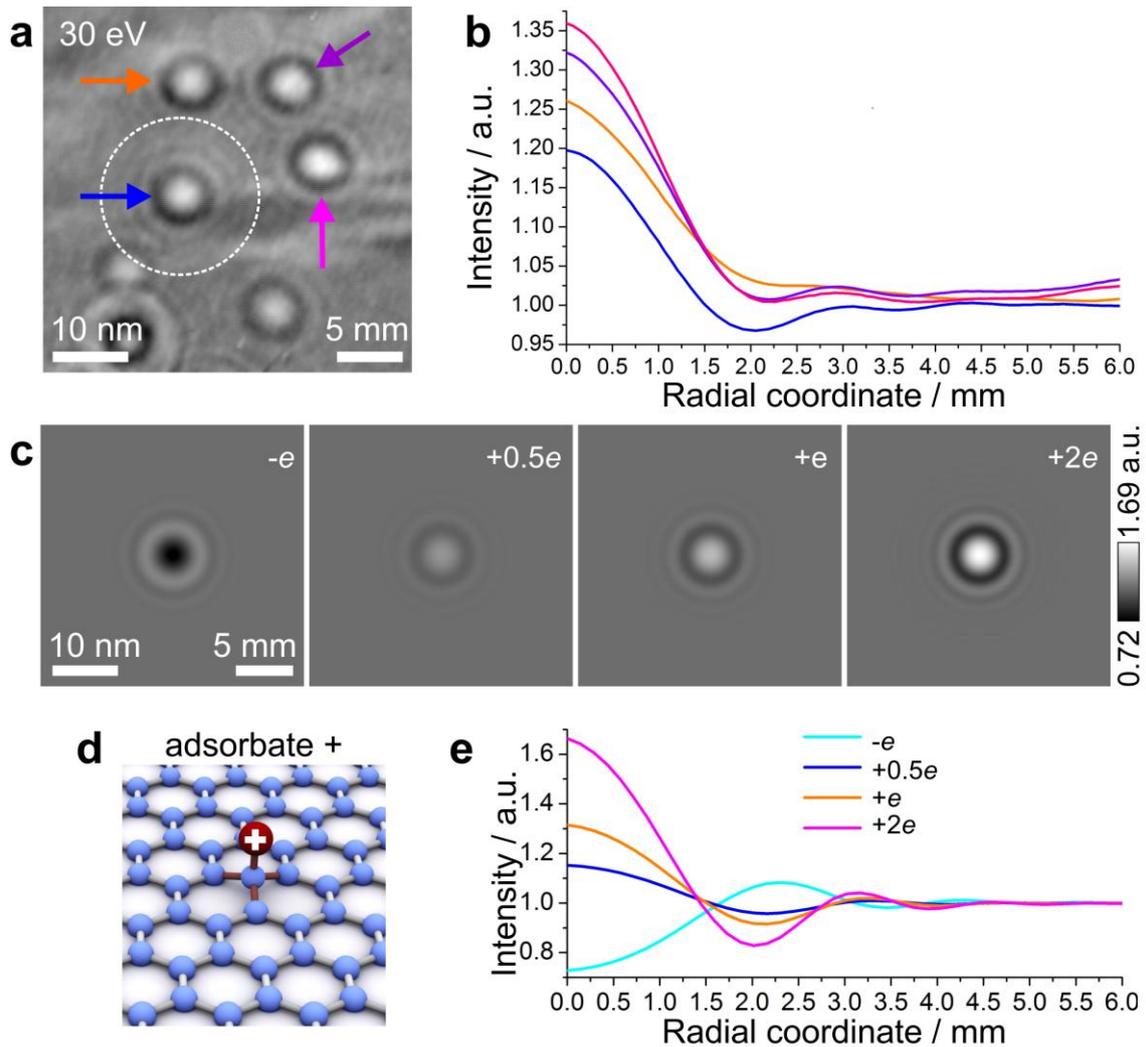

**Figure 2.** Holograms of charged adsorbates. (a) A hologram exhibiting bright spots, recorded with 30 eV electrons, at a source-to-detector distance of 47 mm and a source-to-sample distance of 82 nm, at a resolution of 0.6 nm. (b) Intensity profiles corresponding to the four bright spots marked in (a) averaged over the angular coordinate as a function of the radial coordinate counted from the centre of the spot. The intensity distribution range extends out to 6 mm on the detector corresponding to the radius of the dashed circle indicated in (a). (c) Simulated holograms of a point charge of four different charge values at an electron energy of 30 eV at a source-to-detector distance of 47 mm and a source-to-sample distance of 82 nm. (d) Schematic representation of a charged adsorbate on graphene. (e) The angular averaged intensity profiles as a function of the radial coordinate calculated from the simulated holograms shown in (c). The scale bars in (a) and (c) indicate the sizes in the object plane (left) and in the detector plane (right).

## 3.1 Charges of opposite sign

In addition to the spots discussed above, there are spots exhibiting non-rotational symmetric contrast, as apparent for example in the hologram in Figure 3a. The observed contrast characteristics are in compliance with the ones arising in simulations concerning two oppositely charged adsorbates in close proximity as illustrated in Figure 3c. Note that the spot in the right bottom corner in Figure 3a suddenly turns into a regular bright spot, indicating that the negative charge has vanished, while the positive charge is still present. For this very spot, the distance between the two charges, estimated from the distance between the centers of the two sets of the concentric rings, as shown in Figure 3b, amounts to $1.4 \pm 0.8$ nm. The simulated holograms of two charges, $+0.7e$ and $-0.7e$, separated by 0.5 nm, 1 nm, 1.5 nm and 2 nm are shown in Figure 3c. From the fitting of the radial intensity curve of the hologram of the remaining positive charge, as illustrated in Figure 3d, we estimate the charge to be $(+0.7 \pm 0.2)e$. The comparison with the profiles of the simulated holograms of two separated charges, shown in Figure 3e, leads to a distance of $(1.5 \pm 0.5)$ nm between the charges.

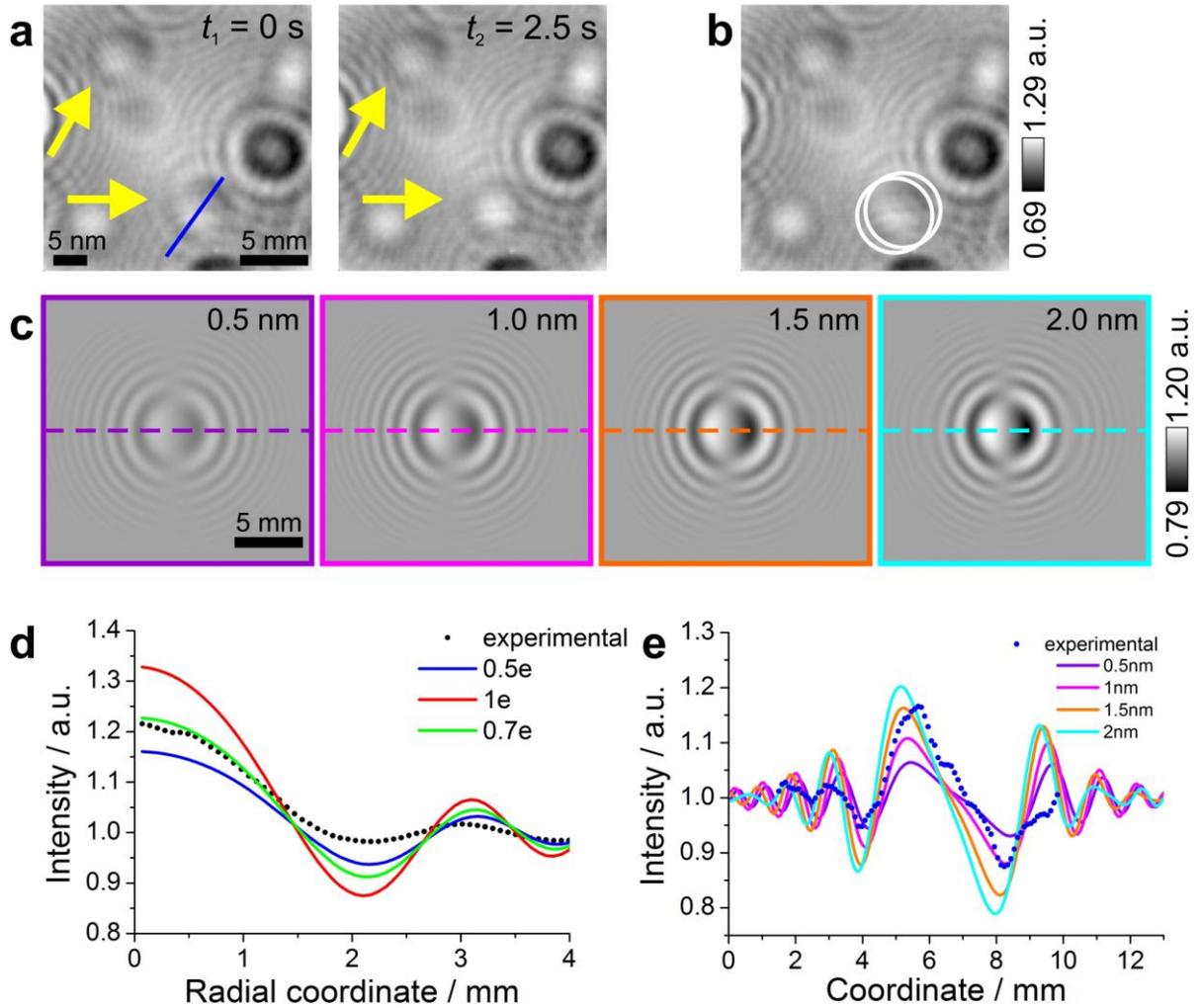

**Figure 3.** Charges of opposite sign. (a) Experimental hologram recorded with 30 eV electrons, at a source-to-detector distance of 47 mm and a source-to-sample distance of 95 nm, at a resolution of 0.5 nm. Two spots with gradient contrast, attributed to a positive charge on the left and a negative charge on the right side, are indicated by the yellow arrows, and the hologram acquired 2.5 s later is showing that the spot of former gradient contrast at the bottom right has turned into a bright spot. (b) The same hologram as in (a, left), but with white circles to denote the difference in position of the positive and negative charges. (c) Simulated holograms of two charges, $+0.7e$ (left) and $-0.7e$ (right) separated by 0.5 nm, 1.0 nm, 1.5 nm and 2.0 nm. In the simulations the source-to-detector distance is 47 mm and the source-to-sample distance is 95 nm. (d) The averaged radial intensity profiles of the bright spot at the bottom right in the experimental hologram (a, right) and that of the related simulated holograms of a positive charge of $0.5e$, $07e$ and $1e$. (e) The intensity profiles through the experimental (a, left) and simulated (c) holograms along the coloured lines.

## 4. Intensity dynamics: blinking and associated charge reversal

Under continuous observation by low-energy electrons, the bright spots exhibit intensity dynamics: blinking, complete disappearance as well as reversal of the contrast from bright to dark and inversely. We observed such intensity dynamics at different electron energies, ranging from 30 eV to 129 eV and currents between 10 nA and 500 nA. An example is shown in Figure 4 (Movie S2) where the intensity maxima for all three bright spots marked in the hologram in Figure 4a, show the same value of about 1.19 a.u., see plot in Figure 4b. The spots also exhibit an intermediate intensity at about 1.0 a.u. that we define as neutral state; at this intensity the spots can no longer be distinguished from the background. One of the spots (see spot 1 in Figure 4b) displays an inversion of its contrast from bright to dark with a minimum of the intensity at about 0.9 a.u., as shown in Figure 4c at the frame $t = 180$ s. The experimental observations of features changing their contrast from bright to dark can be attributed to a change in the charged state, while a vanishing contrast is attributed to a neutral state of an adsorbate too small to be resolved. These transitions between the states can either be caused by a charge re-distribution between adsorbate and graphene, by the impact of an imaging electron or by phonon scattering. We estimate an electron current density of $1.56 \cdot 10^6$ $e/(s \cdot nm^2)$ for a total emission current of 10 nA. This amounts to about $2.40 \cdot 10^4$ $e/s$ for a circle with a radius of 70 pm occupied by a carbon atom. However, we found no correlation between the variation of the intensity of the illuminating electron beam and the dynamics of the change in the charge state for a selected spot, as discussed in the Supporting Information.

The charged adsorbates exhibit different intensity fluctuation characteristics. Some adsorbates remain positively charged over long time with occasional transitions to neutral and negatively charged states while others turn into a stable negatively charged state. The intensity of one of the spots, marked by a blue circle in Figure 4d, has been investigated in more detail. 15083 frames were acquired at a frame rate of 60 fps. The intensity of the spot as a function of the frame number is shown in Figure 4e. The total number of frames in each of the three states are: $N_{-1} = 985$, $N_0 = 1310$ and $N_{+1} = 12788$, and the related total time durations in each state are: $T_{-1} = 16.42$ s, $T_0 = 21.83$ s and $T_{+1} = 213.13$ s, where −1, 0 and +1 are negatively charged, neutral and positively charged states, accordingly. This translates into the probabilities of finding the adsorbate in a selected state: $p_{-1} = 0.065$, $p_0 = 0.087$ and $p_{+1} = 0.848$. To get a first, albeit still rough, idea about the energetics involved, in using the formula for the ratio of probabilities $\frac{p_i}{p_j} = \exp\left(\left(E_j - E_i\right)/k_B T\right)$, where $E_i$ is the energy of state $i$, and $k_B$ denotes the Boltzmann constant, we estimate the free energy differences for the three states: $E_0 - E_{+1} = 59$ meV and $E_{-1} - E_0 = 7$ meV.

This however assumes that the population of each state is representative for the distribution under ordinary thermal equilibrium, not affected by the kinetics of the transitions. To be on the safe side, we also computed the population distribution by taking only every 50$^{th}$ frame into account. In this way we ensure that the system has had a chance to frequently change its state during this elongated time interval before every 50$^{th}$ frame is used to probe its configuration. This analysis leads to: $N_{-1} = 21$, $N_0 = 27$ and $N_{+1} = 254$ with the corresponding probabilities: $p_{-1} = 0.069$, $p_0 = 0.089$ and $p_{+1} = 0.841$ leading to: $E_0 - E_{+1} = 58 \pm 7$ meV and $E_{-1} - E_0 = 6 \pm 2$ meV, values that are comparable to the ones above within the statistical errors given by the square root of the number of observations. However, these values should so far only be taken as an order of magnitude of the energetics involved.

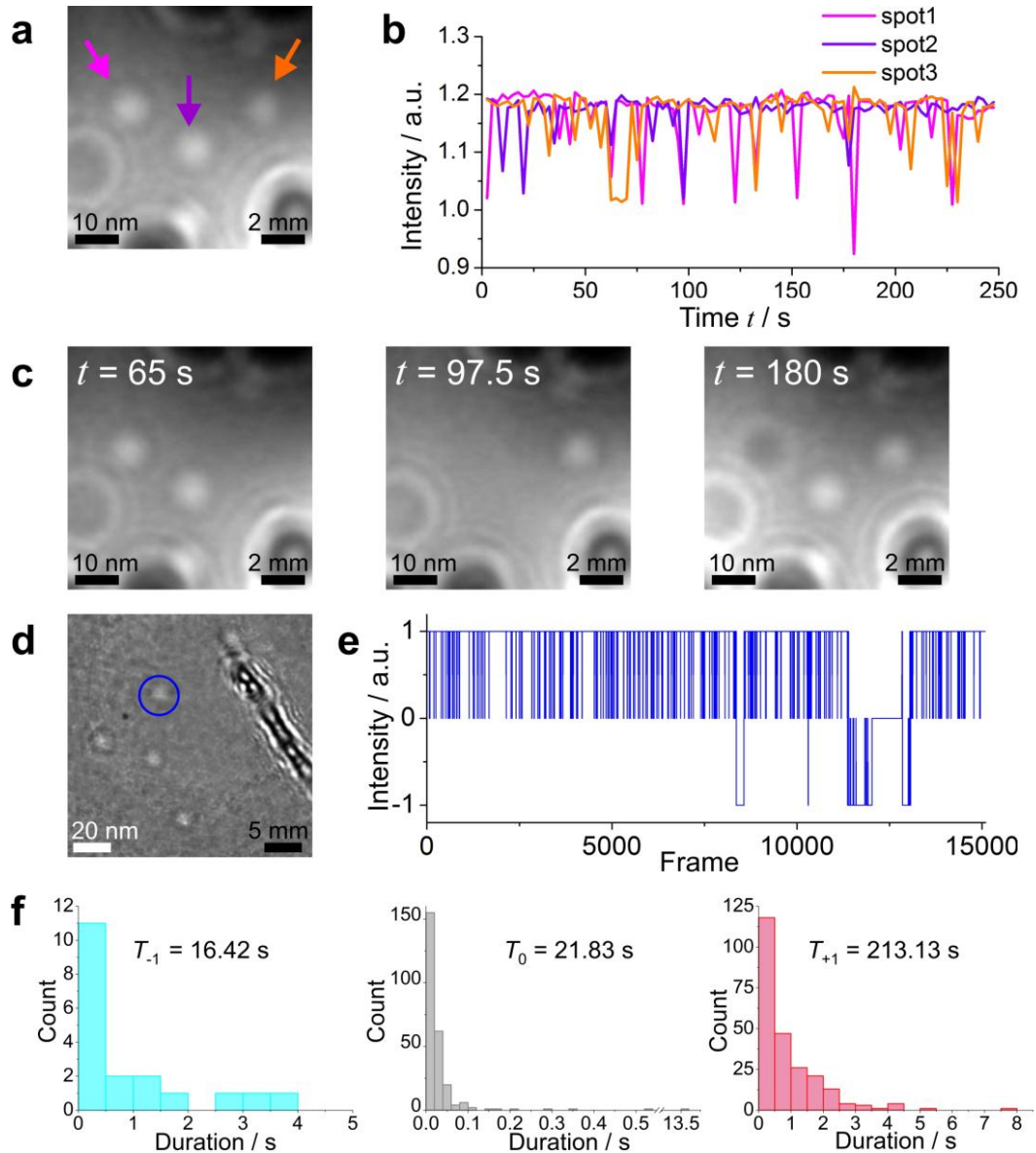

**Figure 4.** Intensity dynamics of bright features in holograms (Movie S2). (a) A hologram exhibiting spots acquired with 38 eV energy electrons at a source-to-detector distance of 47 mm and a source-to-sample distance of 235 nm, at a resolution of 1.7 nm. (b) Maximal intensity as a function of time for each of the three bright spots marked in (a). A total of 99 frames were acquired during 247.5 s. (c) Region of interest recorded at different times. (d) A selected region in a hologram acquired with 129 eV energy at a source-to-detector distance of 70 mm and a source-to-sample distance of 280 nm, at a resolution of 1.3 nm. (e) Relative intensity at the spot shown in the blue circle in (d) as a function of frame number. (f) Three histograms showing the counts for negative, neutral and positive states of the selected spot. The scale bars in (a), (c) and (d) indicate the sizes in the object plane (left) and in the detector plane (right).

## 5. Mobility of charged adsorbates

In the following we take advantage of the enhanced holographic contrast of a charged object that enables tracking of entities too small to be detected otherwise. Adatoms are adsorbed at distinct sites on graphene, and they can migrate until they find an energetically favourable site[26]. DFT calculations show that most chemical elements, including Al, Si, Pt, Pd, Au, Cu and others have a diffusion barrier energy less than 0.4 eV and several elements such as V, Cr, Mn, Fe Co, Mo and Ru have a diffusion barrier energy higher than 0.4 eV[27]. The diffusion barrier energies of Au, Cr, and Al adatoms on pristine monolayer, bilayer and trilayer graphene were found to be of the order of $k_\text{B}T$ at room temperature (25.7 meV) or even smaller[28]. This implies that adatoms of most chemical elements can easily and quickly migrate across the graphene lattice before they bind to an energetically favourable site. Hardcastle et al[28] performed scanning transmission electron microscopy (STEM) on graphene samples and they speculated that all metal adatoms are highly mobile on graphene, but that they all migrated to stable sites before the samples were characterized in the microscope. We also observed that the positively charged adsorbates, represented by bright spots in the hologram, exhibit some mobility on the graphene but only in the first few seconds during the exposure to the low-energy electron beam. Afterwards no movement of the bright spots was observed. However, the negatively charged adsorbates, represented by dark spots in the hologram, often carry out a random walk during the electron exposure, as for example illustrated in Figure 5. For a selected adsorbate shown in Figure 5a, its random walk, shown in Figure 5b, is characterized by the mean squared displacement: $\overline{r^2} = \frac{1}{N} \sum_{i=1}^{N-1} \left[ (x_{i+1} - x_i)^2 + (y_{i+1} - y_i)^2 \right] = (3.33 \pm 0.17)$ nm$^2$, where $N$ is the number of frames $N = 271$ acquired at a time interval $\Delta t = 1/60$ s $= 0.0167$ s for a total time of 4.5 s, and $(x_i, y_i)$ are the adsorbate coordinates at frame $i$. This leads to a diffusion coefficient of $D = \overline{r^2} / (4\Delta t) = (50.02 \pm 2.61)$ nm$^2$/s.

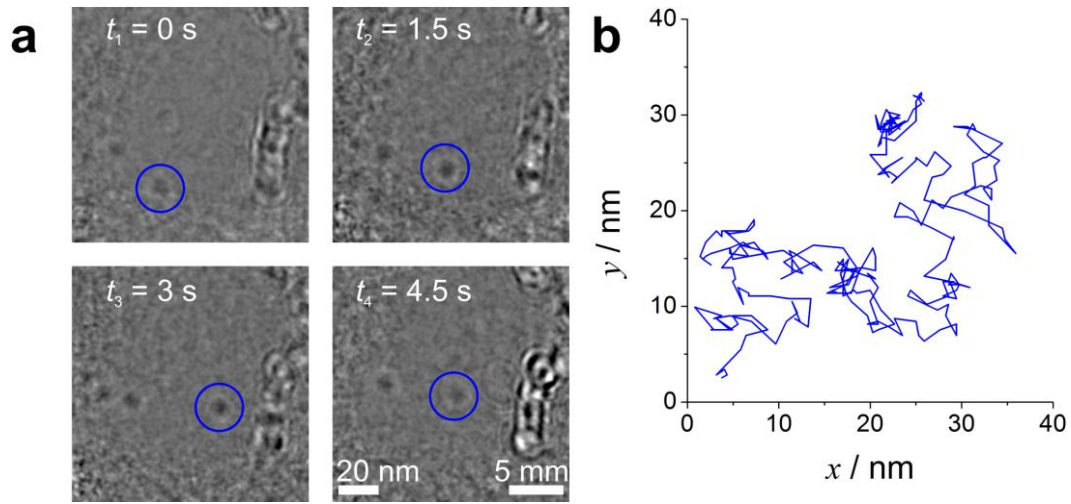

**Figure 5.** Mobility of adsorbates on graphene. Holograms were recorded with 125 eV electrons, at a source-to-detector distance of 70 mm and a source-to-sample distance of 380 nm, at a resolution of 1.7 nm. (a) Four contrast-enhanced holograms from a series of holograms (Movie S3) showing the motion of a negatively charged adsorbate on graphene. The scale bars indicate the sizes in the object plane (left) and in the detector plane (right). (b) Trajectory of the adsorbate marked with blue circles in (a), followed over a duration of 4.5 s.

## 6. Discussion

We have shown that the high sensitivity of low-energy electrons to local electric fields can be employed to directly visualize charge distributions with a sensitivity of a fraction of an elementary charge carried by adsorbates on freestanding graphene and to quantitatively study their behaviour. By means of low-energy electron holography we were able to observe charge transfer processes. While positively charged states are found to be the most frequently observed ones, adsorbates are also occasionally neutralised or are undergoing a transition to a negatively charged state with a comparably short life-time. Pairs of two oppositely charged adsorbates separated by a distance in the order of one nanometer were also found, and for a selected pair of such charges the estimated distance amounts to 1.5 nm. Charged adsorbates, mainly those carrying a negative charge, perform a random walk on freestanding graphene. Although adsorbates are known to quickly occupy some stable sites on graphene, we were able to observe their random walk behaviour on graphene before they were finally trapped. For future development, a more complicated theoretical model which takes into account the charge re-distribution within the atom and also considers the graphene support can be developed to arrive at more precise estimations of the charge values.

**Author Contributions**

T. L. proposed studying the origin of bright spots observed on graphene, suggested an explanation of the effect, performed the related simulations and data analysis. F. W., J.-N. L., C.E. and H.-W. F. developed the low-energy electron holographic microscope used in this study. F. W. and J.-N. L. performed the experiments. T. L., F. W. and H.-W. F. wrote the manuscript in discussions with all remaining authors.

**Funding Sources**

Swiss National Science Foundation, grant numbers PZ00P2_148084 and 200021_150049.

**Acknowledgements**

We would like to thank the Swiss National Science Foundation for financial support (grant numbers PZ00P2_148084 and 200021_150049).

# SUPPORTING INFORMATION

# Contents

# 1. Resolution estimation

The resolution with which an object is reconstructed from its hologram can be estimated from the Fourier spectrum of the hologram or the reconstructed object[1]. We define the maximal frequency in the spectrum $k_{max}$ as the frequency where the peaks of the spectrum are still distinguishable from noise, as illustrated in Figure S1. The resolution is then given by: $R = \frac{2\pi}{k_{max}}$ The spectrum of the hologram shown in Figure 1b is displayed in Figure S1; with values $k_{max} = 6.15$ nm$^{-1}$ and $R = 1.0$ nm.

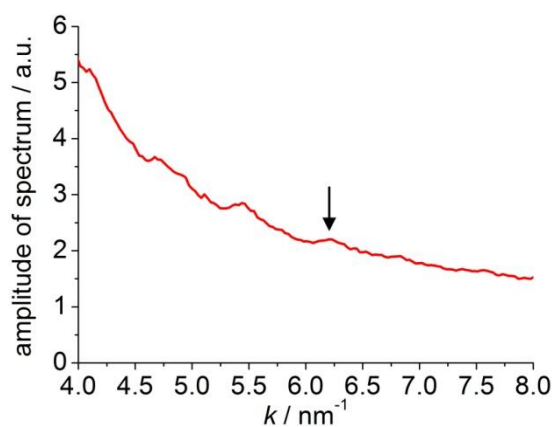

**Figure S1. Resolution estimation.** Radial distribution of the amplitude of the Fourier spectrum of a hologram $\mathrm{FT}(H(X,Y))$ as function of the Fourier domain coordinate $k$. The arrow indicates $k_{max}$.

## 2. Calculation of the transmission function for simulating the hologram of a point charge

The electric potential distribution of a point-like charge $q$ in free space is given by:

$$\varphi(r) = \frac{q}{4\pi\varepsilon_0} \frac{1}{r} \tag{S1}$$

and the corresponding electric field distribution is found as:

$$\vec{E}(r) = -\mathrm{grad}\,\varphi(r) = \frac{q}{4\pi\varepsilon_0} \frac{1}{r^2} \vec{e}_r, \tag{S2}$$

where $\varepsilon_0$ is the vacuum permittivity, $r$ is the distance from the charge and $\vec{e}_r$ is the unit vector in the spherical coordinate system.

The movement of an electron is described by the equation:

$$m\ddot{\vec{r}} = -e\vec{E}(r), \tag{S3}$$

where $e$ is the elementary charge and $m$ is the mass of the electron.

We are only interested in the $(x, z)$ projection due to the rotational symmetry around the $z$ axis. We consider the electron moving parallel to the $z$-axis at a speed $v_0$.

For infinitesimal small $\Delta x$ and $\Delta z$, equations for $x$ and $z$ coordinates are:

$$\frac{\Delta v_x}{\Delta t} = -\frac{e}{m} E_x(x, y, z)$$
$$\frac{\Delta v_z}{\Delta t} = -\frac{e}{m} E_z(x, y, z), \tag{S4}$$

or

$$\Delta v_x = -\frac{e}{m} E_x(x, y, z)\Delta t$$
$$\Delta v_z = -\frac{e}{m} E_z(x, y, z)\Delta t. \tag{S5}$$

Since the velocity in the $x$-direction can be expressed as:

$$v_x = v_{ox} + \Delta v_x \tag{S6}$$

where $v_{ox} = 0$, we can re-write Eq. S5:

$$v_x = -\frac{e}{m} E_x(x, y, z)\Delta t$$
$$v_z = -\frac{e}{m} E_z(x, y, z)\Delta t + v_0 \approx v_0, \tag{S7}$$

where the approximation in Eq. S7 can be applied since at an electron energy of 30 eV the following expression holds: $\max\left|\frac{e}{m}E_z(x,y,z)\Delta t\right| \approx 0.022 v_0$. By solving the last equations for coordinates, we obtain:

$$\frac{\Delta x}{\Delta t} = -\frac{e}{m}E_x(x,y,z)\Delta t$$
$$\frac{\Delta z}{\Delta t} \approx v_0. \tag{S8}$$

or

$$\Delta x = -\frac{e}{m}E_x(x,y,z)(\Delta t)^2$$
$$\Delta t \approx \frac{\Delta z}{v_0}. \tag{S9}$$

By substituting $\Delta t$ into the expression for $\Delta x$, we obtain:

$$\Delta x = -\frac{e}{m}E_x(x,y,z)\left(\frac{\Delta z}{v_0}\right)^2. \tag{S10}$$

The deflection of an electron passing the electric field can then be expressed as:

$$\gamma(x) = \frac{\Delta x}{\Delta z} = -\frac{e}{mv_0^2}E_x(x,y,z)\Delta z, \tag{S11}$$

$E_x(x,y,z)$ is calculated as:

$$E_x(x,y,z) = \frac{q}{4\pi\varepsilon_0}\frac{x}{\left(x^2+y^2+z^2\right)^{3/2}}. \tag{S12}$$

Next, we simplify the z-dependency of the electric field distribution by replacing it with the distribution at the fixed z-coordinate where the point-like charge is located ($z = 0$):

$$E_x(x,y,z) = \frac{q}{4\pi\varepsilon_0}\frac{1}{x^2}. \tag{S13}$$

By substituting Eq. S13 into Eq. S11, we obtain:

$$\tan(\gamma(x)) = -\frac{e}{mv_0^2}\frac{q}{4\pi\varepsilon_0}\frac{1}{x^2}\Delta z. \tag{S14}$$

The deflection of an electron is maximal when it passes the charge in close proximity. Therefore we consider a distance $a$ of a few Angstrom as a reasonable approximation for the interaction region, see Figure S2. According to Eq. S14 the less the energy of an electron, the larger is the deflection angle, which can be explained that a slower electron spends more time in the electric field of the charge.

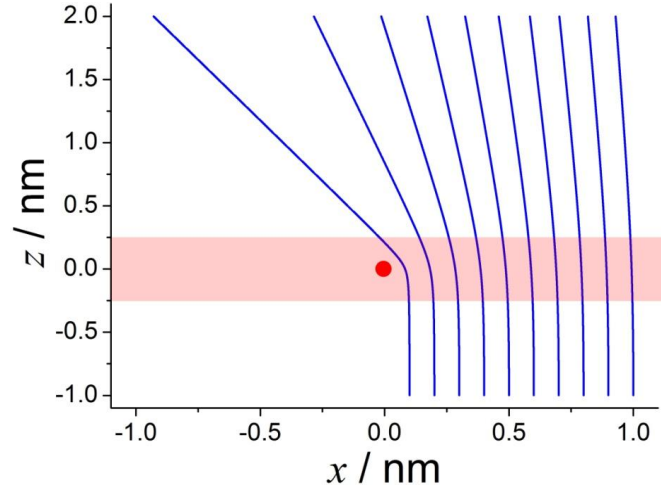

**Figure S2. Simulated trajectories for 30 eV electrons passing a positive charge $e$.** The region marked exhibits a height of $2a = 5$Å.

We set $\Delta z = 2a$, which provides

$$\gamma(x) = \alpha(x)\arctan\left(\frac{e}{mv_0^2}\frac{q}{4\pi\varepsilon_0}\frac{2a}{x^2}\right), \quad (S15)$$

whereby $\alpha(x) = 1$ for $x < 0$ and $\alpha(x) = -1$ for $x > 0$. The total phase shift introduced into an electron wave can be represented as a phase shift in the object plane:

$$\Delta\chi(x) = -\frac{2\pi}{\lambda}|\gamma(x)||x| \approx -\frac{2\pi}{\lambda}\frac{e}{mv_0^2}\frac{q}{4\pi\varepsilon_0}\frac{2a}{|x|} \quad (S16)$$

We set $2a = 5$Å as shown in Figure S2, as the region where significant bending of the electron trajectories occurs.

The total transmission function in the object plane is then given by:

$$T(x, y) = \exp(-A(x, y))\exp(i\Delta\chi(x, y)), \quad (S17)$$

whereby $A(x, y)$ is the absorption distribution. For a single point-like charge we assume zero absorption and thus $\exp(-A(x, y)) = 1$.

## 3. Simulated holograms of holes in and adatoms on graphene

### 3.1 Holes

A hole, a prominent defect in graphene, is illustrated in Figure S3a. It was simulated as a fully transparent region in a sheet exhibiting a complex-valued transmission function which takes scattering off carbon atoms into account:

$$T(x,y) = 1 - g(x_i, y_i) + \alpha g(x_i, y_i) f(\vartheta) e^{i\varphi(\vartheta)}, \quad (S18)$$

where $g(x_i, y_i)$ is 1 at the position $(x_i, y_i)$ of carbon atom $i$ and 0 elsewhere, $\alpha$ represents the fraction of elastically scattered electrons, $f(\vartheta)$ is the amplitude of the scattered wave and $\varphi(\vartheta)$ is the phase of the scattered wave. The phase shifts for the simulation are taken from the NIST library and for 50 eV electrons we obtain: $f(\vartheta = 0) = 7.389$ and $\varphi(\vartheta = 0) = 0.725$ rad.

In the absence of the graphene patch, $T(x,y) = 1$, and a wave which passes through the aperture is described in the detector plane as $U_0(X,Y)$.

With graphene being present, the incoming wave is partly absorbed by graphene

$$T(x,y) = 1 - g(x_i, y_i) \quad (S19)$$

where $G(x_i, y_i)$ is 1 at the position $(x_i, y_i)$ of carbon atom $i$ and 0 elsewhere. A wave passing through an aperture in graphene is thus described in the detector plane as $U_0(X,Y) - U_G(X,Y)$.

By taking the absorption as well as the forward scattering of carbon atoms into account, we obtain:

$$T(x,y) \approx 1 - g(x_i, y_i) + \alpha g(x_i, y_i) f(0) e^{i\varphi(0)}, \quad (S20)$$

and the wave in the detector plane is approximately described by

$$U_0(X,Y) - U_G(X,Y) + \alpha G(X,Y) f(0) e^{i\varphi(0)}, \quad (S21)$$

where $f(0)$ denotes the amplitude and $\varphi(0)$ the phase of the scattered wave in forward direction, and $G(X,Y)$ is the complex-valued distribution describing the wave scattered by graphene described by $g(x_i, y_i)$.

Since a transmission of 73% was measured for low-energy electrons through graphene[2], we simulated the situation with 50 eV electrons passing through a 40 nm in diameter patch of graphene. Under such condition, we found that the intensity of the transmitted wave amounts to 73% of the initial intensity when $\alpha = 0.073$. With these parameters, the transmission function of graphene given by Eq. S18 was simplified to $T(x, y) = 0.856e^{0.105i}$. Thus, in the simulation graphene is described as a sheet exhibiting the complex-valued transmission function $T(x, y) = 0.856e^{0.105i}$ containing a hole. The algorithm for the simulation of hologram formation with spherical wave is explained in detail elsewhere[3].

Figure S3b shows the central 200 × 200 pixels region of the simulated holograms for holes of 0.5, 3 and 5 nm in diameter. The holograms are normalized by division with the background formed by the wave passing through a defect-free graphene sheet. Figure S3c shows the corresponding radial intensity profiles. The following conclusions can be drawn from the data shown in Figure S3b – c: (1) For any hole exhibiting a diameter of a few times the wavelength there is a decrease of intensity in the centre of the hologram. Such decrease was experimentally not observed in the centre of the bright spots. (2) The holograms of holes with larger diameter exhibit a relatively high intensity in the centre displaying values close to the experimentally observed intensity. (3) The holograms of holes with larger diameter show a pronounced first minimum and its radial position is close to the radial position of the first minimum observed within the bright features of the experimental holograms.

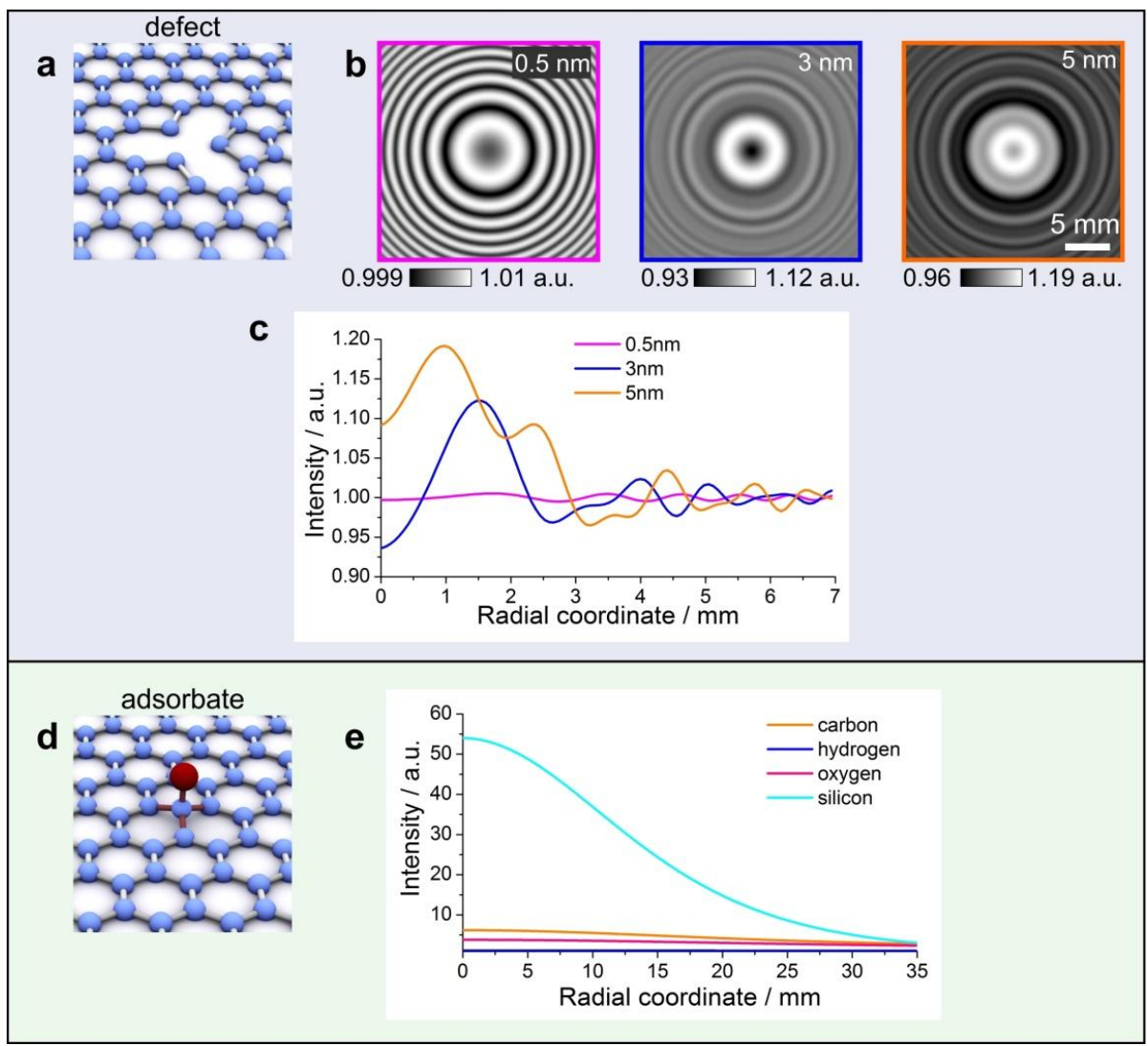

**Figure S3. Simulated holograms of holes in graphene and adatoms on graphene. a**, Artistic representation of a hole in graphene. **b**, Simulated holograms of holes of different diameter in a graphene sheet. **c**, Angular averaged intensity profiles as a function of the radial coordinate calculated from the simulated holograms shown in **b**. **d**, Artistic representation of a single atom adsorbate on graphene. **e**, Angular averaged intensity profiles as a function of the radial coordinate calculated from the distribution of the intensity of the 50 eV electron wave scattered off individual atoms. In the simulations, the electron energy amounts to 50 eV, the source-to-detector distance to 47 mm and the source-to-sample distance to 82 nm.

## 3.2 Single atoms on graphene

Here, we simulate a situation where a coherent low-energy electron wave is scattered off a single atom on graphene, as illustrated in Figure S3d by accounting for the anisotropic scattering typical for low-energy electrons. We selected a few adsorbate atoms which can typically be found on graphene: carbon, hydrogen, oxygen and silicon. The complex-valued amplitudes of the scattered electron wave were constructed using the partial wave expansion[4-5]:

$$f(\vartheta) = \sum_{l=0}^{\infty}(2l+1)\frac{1}{k}e^{i\delta_l(k)}\sin\delta_l(k)P_l(\cos\vartheta), \qquad (S22)$$

where $k$ is the wave number, $P_l(\cos\vartheta)$ are Legendre polynomials, $\vartheta$ is the scattering angle, $l$ is the angular momentum number for each partial wave ($l=0$ corresponds to isotropic s-waves, and so on), and $\delta_l(k)$ are the phase shifts. The complex-valued scattering amplitudes were calculated using the phase shifts $\delta_l(k)$ provided by the NIST library[6].

The electron energy was selected to be 50 eV, the lowest energy for which the NIST library provides the phase shifts and representative for the experimental energy range. The simulation resulted in the complex-valued wave $U_0$ originating from a point-like source. To create a more realistic distribution for a single atom a Gaussian-like distribution was imposed such that its full width at half maxima equals twice the empiric covalent radii (70 pm for carbon, 25 pm for hydrogen, 60 pm for oxygen and 110 pm for silicon). The complex-valued wave at the detector was obtained by a convolution of $U_0$ with the atom distribution. Next, a reference wave with an amplitude equal to the maximum of the amplitude of a wave scattered by a carbon atom was superimposed and the hologram computed as the squared absolute value of the result. It is evident that for all waves scattered off individual atoms, the intensity distribution is broad, not exhibiting a pronounced peak within the detector area. The scattering off an individual atom thus contributes to the featureless background only, while a pronounced peak can only be observed for a charged adsorbate. Thus, we can exclude that a single atom is the cause for the bright spots observed in the experimental holograms.

# 4. Simulated holograms of charges at different electron energies

In the simulated holograms of charges at different electron energies shown in Figure S4a – b, a positive charge results in a bright spot and a negative charge results in a dark spot and this contrast is preserved at all electron energies. It can be seen from Figure S4c that the positions of the first minima are found at smaller radial coordinates when the energy of the electrons is increased. A higher electron energy and thus shorter wavelength leads to a downscaled pattern, as also apparent in the holograms in Figure S4a – b. It should also be noted that the amplitude of the zero-order diffraction spot varies with the electron wavelength as well. Thus, for a quantitative estimation of the charge value, the precise energy of the imaging electrons must be known.

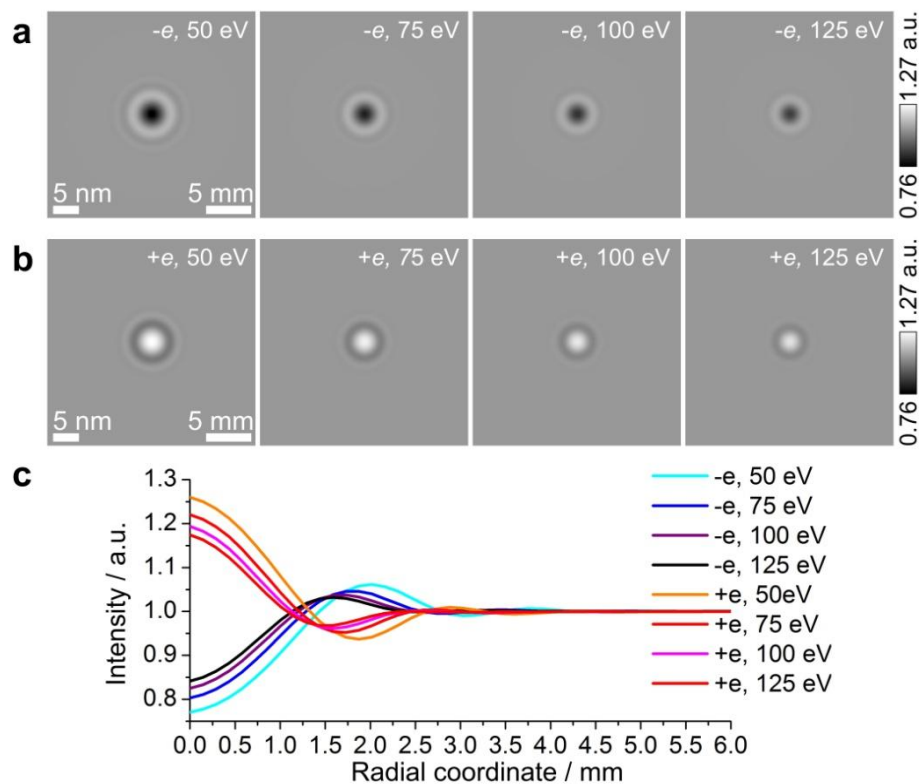

**Figure S4. Simulated holograms of charges at different energies. a**, Holograms of a negative charge. **b**, Holograms of a positive charge. **c**, Angular averaged intensity profiles as a function of the radial coordinate. In all simulations, in order to match the experimental conditions, we set the source-to-sample distance to 82 nm and the source-to-detector distance to 47 mm. The scale bars in (a) and (b) indicate the sizes in the object plane (left) and in the detector plane (right).

# 5. Cross-correlation between the electron beam intensity and the contrast of a selected spot

To evaluate the cross-correlation between the contrast of a selected spot at the detector and the intensity of the electron beam, we estimated the averaged intensity at a region nearby the spot, as shown in Figure S6a. The intensity of the illuminating electron beam as a function of frame number for the first 500 frames is shown in Figure S6b. It displays a mean value of 96.0 a.u with a standard deviation of 4.9 a.u.. The time evolution of the contrast of a selected spot, shown in Figure S6c. The spot is observed bright most of the time corresponding to a positive charge with a few transitions to the neutral state. The cross-correlation between the intensity of the illuminating beam and the contrast of the selected spot is shown in Figure S6d, and indicates no correlation between the two signals.

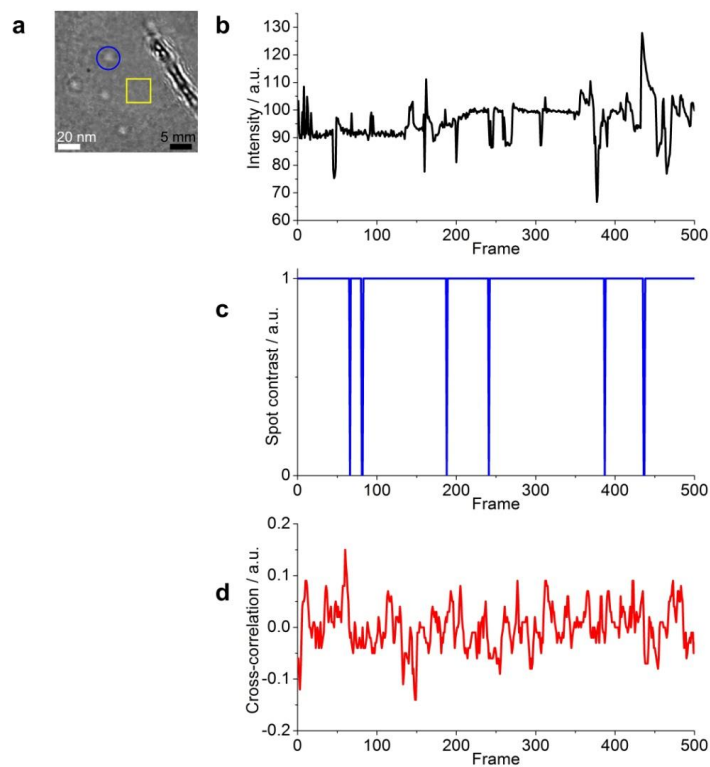

**Figure S6. Cross-correlation between the electron bean intensity and the contrast of a selected spot. a**, A selected spot in the hologram indicated by the blue circle. The selected region for to evaluate beam intensity fluctuations is indicated by the yellow square. Frames were acquired with 129 eV energy at a source-to-detector distance of 70 mm and the source-to-sample distance of 280 nm. The scale bars in (a) indicate the sizes in the object plane (left) and in the detector plane (right). **b**, Intensity as a function of frame number. **c**, Contrast of a selected spot as a function of frame number. **d**, Cross-correlation between the signals shown in (a) and (b).